# Tuning electronic properties in graphene quantum dots by chemical functionalization: Density functional theory calculations

**Hazem Abdelsalam[1], Hanan Elhaes[2], Medhat A. Ibrahim[3]**

[1]Department of Theoretical Physics, National Research Center, Giza, 12622, Egypt

[2]Physics Department, Faculty of Women for Arts, Science, and Education, Ain Shams University, 11757, Cairo, Egypt

[3]Spectroscopy Department, National Research Centre, Giza, 12622, Egypt

The electronic energy gap and total dipole moment of chemically functionalized hexagonal and triangular graphene quantum dots are investigated by the density functional theory. It has been found that the energy gap can be efficiently tuned in the selected clusters by edge passivation with different elements or groups. Edge passivation with oxygen provides a considerable decrease of the large energy gap observed in hexagonal nanodots. The edge states and energy gap in triangular graphene quantum dots can also be manipulated by passivation with fluorine. The total dipole moment strongly depends on: (a) the shape and edge termination of the graphene quantum dot, (b) the attached group, and (c) the position to which the groups are attached. With respect to the shape, edge termination, and the attached group the chemically modified hexagonal-armchair quantum dot has the highest total dipole moment. Depending on the position of the attached groups, the total dipole can be increased, decreased, or eliminated. The significant features, the tunable energy gap and total dipole moment, of the functionalized graphene quantum dots are confirmed by the stability calculations. The obtained positive binding energy and positive frequencies in the infrared spectra imply that all the selected clusters are stable under edge functionalization and passivation with various groups and elements.

# I. INTRODUCTION

Graphene, the truly two dimensional (2D) atomically thin structure of carbon atoms arranged in honeycomb structure, was isolated by mechanical exfoliation in 2004.[1] It has a unique electronic band structure characterized by the massless Dirac spectrum which results in distinct physical properties.[2-5] Due to its fascinating physical properties, graphene is encountered in a variety of applications in the field of electronic, spintronic, photonic, and optoelectronic.[6-11] However, graphene has a zero energy gap which impeded its applications as a semiconductor device.[12, 13] Graphene quantum dots (GQDs), nanoclusters of graphene in which charge carriers are confined in all three directions, is a potential candidate that can extend the applications of graphene as semiconductor material.[14] Recently there has been a great advance in preparation of GQDs with ultra small sizes and well defined edge terminations by chemical methods.[15, 16]

GQDs have a band gap that can be controlled by changing their size, shape, edge morphology, number of layer, and applied external electric and magnetic fields.[17-25] In addition to the tunable band gap, GQDs have discreet energy levels which can be used as qubits in quantum computers.[27, 28] Moreover, they have various edge terminations that can be chemically functionalized[28-33] for numerous applications in nanoelectronic, biosensors, supercapacitors, and nanomedicine.[34-36] It is worth noting that chemical modification of other graphene derivatives, such as fullerene, has been also investigated.[37-39] The enhanced electronic properties and biological activity make it a potential applicant for various applications as gas sensor[40-41] and as anti-HIV protease inhibitors.[42, 43]

In this paper, chemical functionalization of hexagonal and triangular GQDs with zigzag and armchair terminations is considered. Different functional groups and elements are attached to the edges of the clusters, namely amide, cyano, isopropyl, nitro, aldehyde, fluorine, and oxygen. In our work on carboxylation of the same graphene flakes the numbers of COOH groups and the attachment sites that give the highest dipole moment have been specified.[44] For hexagonal GQDs with armchair/zigzag termination the number of COOH that maximize the total dipole moment is 4COOH/5COOH attached to the corners of the cluster, respectively. For the triangular with armchair/zigzag the number of COOH is 2COOH/COOH, respectively. Here we replace these carboxyl groups with the above mentioned groups separately to reveal the effect on the energy gap and the total dipole moment. In what follows, we provide structure generation and computational model in section II, the results and discussion are provided in section III, and the conclusion in section IV.

## II.     CLUSTER BUILDING AND COMPUTATIONAL MODEL

Graphene sheet is built using the arbitrary lattice translational vector L which is defined in term of the graphene lattice vectors $a_1$ and $a_2$ as

$$L = na_1 + ma_2.$$

Where n, m are integers and

$$a_1 = \frac{a}{2}\left(3, \sqrt{3}\right), \ a_2 = \frac{a}{2}(3, -\sqrt{3}), \ and \ a = 1.42 A^0 \text{ is the C-C bond length.}$$

Small flakes with the required shape and edge termination can then be cut and isolated from the graphene sheet. Other structures such as graphene nanoribons superlattices can be isolated from the bulk graphene sheet.[45] The effect of chemical modification on these interesting structures will be the topic of our future work. The xyz coordinates of the generated flakes are used to build the Gaussian formatted files for the density functional theory calculations (DFT).

All the obtained GQDs were then fully optimized at the B3LYB/3-21G level of theory.[46-48] The DFT calculations were performed in Gaussian 09[49] by minimizing the total energy without symmetry constraint. The 3-21G basis set was found to be adequate in size by comparing geometric optimization and electronic properties with bigger basis sets such as 6-31G**.[19, 28] We also perform DFT calculations on pyrene ($C_{16}H_{10}$) using 3-21 G and 6-311G** to justify the sufficiency of 3-21G when considering both results accuracy and computational efficiency.[44] The calculated energy gap by 3-21G (6-311G**) basis set is 3.93 eV (3.85 eV), with CPU time equals 10 min (1.36 h). Hence, with respect to the required CPU time and results accuracy, the 3-21G can be consider as a sufficient basis set.

## III. RESULTS AND DISCUSSION

Attachment of different functional groups (amide, cyano, isopropyl, nitro, and aldehyde) to the edges of graphene nanodots having hexagonal and triangular shapes provides significant enhancement of their electronic properties. The abbreviations ATRI and AHEX refer to triangular and hexagonal GQDs with armchair termination, respectively. The ZTRI and ZHEX refer to triangular and hexagonal GQDs with zigzag edges, respectively. The electronic density of states, band gap, charge distribution, and total dipole moment are investigated. Moreover, the frequencies of the infrared spectra are calculated to confirm the stability of the functionalized GQDs.

### A. STRUCTURE STABILITY

Figures 1 and 2 show the optimized structures of the chemically functionalized clusters: AHEX (FIG. 1, a-e), ZHEX (FIG. 1, f-j), ATRI (FIG. 2, a-e), and ZTRI (FIG. 2, f-j). In all the clusters the edge atoms are saturated with hydrogen, the C-H bond lengths are in the range from 1.077 to 1.085 $A^0$. The optimized bond lengths and the binding energies ($E_B$) of all the clusters are presented in TABLE I. The length of the bond between an attached group ($X$) and the corresponding carbon edge atom ($d_{XC}$) depends on the attached group, for instance,

$d_{XC}$ =1.377 $A^0$ for attached 4 amide groups, whereas $d_{XC}$ =1.527 $A^0$ for attached 4 isopropyl groups to the AHEX, as seen in TABLE I. The carbon-carbon (C-C) bond length varies from 1.362 to 1.466 $A^0$ with the short and long bond lengths being observed at the edge and in the bulk of the dot, respectively. The effect of the attached functional groups on the C-C bond length is negligible. The graphene flakes do not preserve their plain structure after functionalization. The bond angles ranges from $118.5^0$ to $122.5^0$ and small buckling lengths (difference between atomic sites in z-direction) equal to $3x10^{-4}$ between adjacent edge atoms and $6x10^{-5}$ $A^0$ between bulk atoms are observed.

TABLE I. The optimized bond lengths between the attached group and the edge atoms ($d_{XC}$), the C-C bond lengths, and the binding energies of the functionalized graphene quantum dots are presented.

| Structure | | $d_{XC}(A^0)$ | $d_{CC}(A^0)$ | $E_B$ (eV) |
|---|---|---|---|---|
| **AHEX** ($C_{42}H_{18}$) | 4 amide | 1.377 | 1.388-1.470 | 1.8853 |
| | 4 cyano | 1.426 | 1.387-1.448 | 2.0038 |
| | 4 isopropyl | 1.527 | 1.391-1.466 | 1.3530 |
| | 4 nitro | 1.464 | 1.381-1.461 | 1.8787 |
| | 4 aldehyde | 1.478, 1.479 | 1.387-1.464 | 1.8730 |
| **ZHEX** ($C_{54}H_{18}$) | 5 amide | 1.379-1.388 | 1.363-1.461 | 1.9569 |
| | 5 cyano | 1.424 | 1.372-1.454 | 2.0811 |
| | 5 isopropyl | 1.530-1.534 | 1.366-1.460 | 1.3941 |
| | 5 nitro | 1.473-1.477 | 1.362-1.450 | 1.9542 |
| | 5 aldehyde | 1.477-1.486 | 1.369-1.454 | 1.9416 |
| **ATRI** ($C_{60}H_{24}$) | 2 amide | 1.374 | 1.372-1.462 | 0.7276 |
| | 2 cyano | 1.425 | 1.372-1.460 | 0.7438 |
| | 2 isopropyl | 1.525 | 1.372-1.459 | 0.6369 |
| | 2 nitro | 1.460 | 1.371-1.459 | 0.7250 |
| | 2 aldehyde | 1.476, 1.475 | 1.372-1.459 | 0.7226 |
| **ZTRI** ($C_{46}H_{18}$) | Amide | 1.374 | 1.392-1.443 | 0.4836 |
| | Cyano | 1.468 | 1.404-1.443 | 5.7267 |
| | isopropyl | 1.528 | 1.392-1.449 | 0.4437 |
| | Nitro | 1.471 | 1.392-1.444 | 0.4819 |
| | Aldehyde | 1.480 | 1.393-1.446 | 0.4804 |

The stability of the studied clusters is confirmed by the positive values of the binding energy, which is calculated as:

$E_B = (E_C - (nE_X + E_{GQDs}))/N$,

where $E_C$ is the compound (GQDs + the functional groups) total energy, $E_X$ and $E_{GQDs}$ is the functional group and the GQDs total energies, respectively. N is the total number of atoms and n is the number of attached groups. As can be seen from TABLE I, the binding energies of cyano group to different GQDs are higher than those for all other groups.

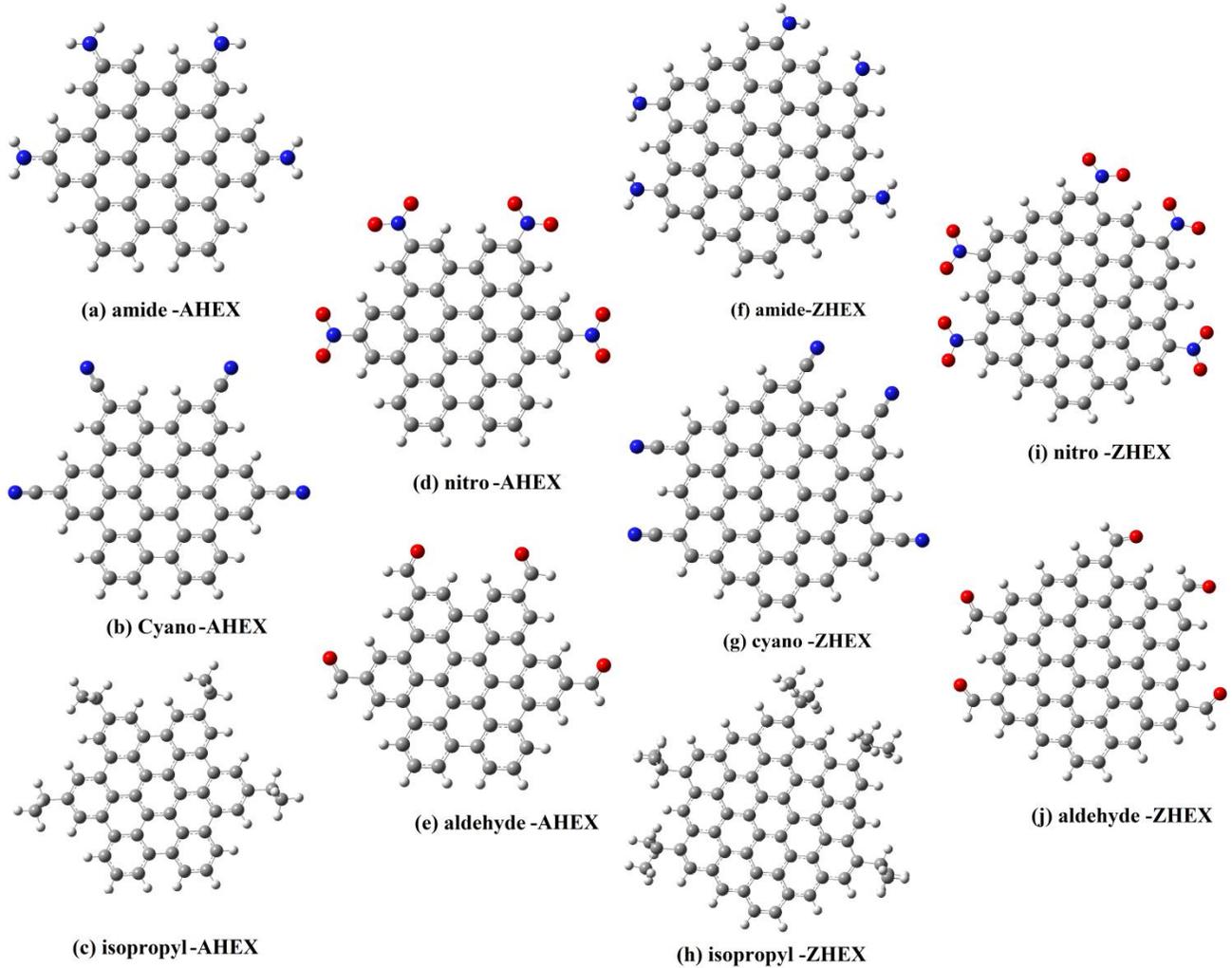

FIG. 1. (a-e) the optimized structures of chemically functionalized AHEX with different groups (f-j) the ZHEX optimized structures attached to the same functional groups.

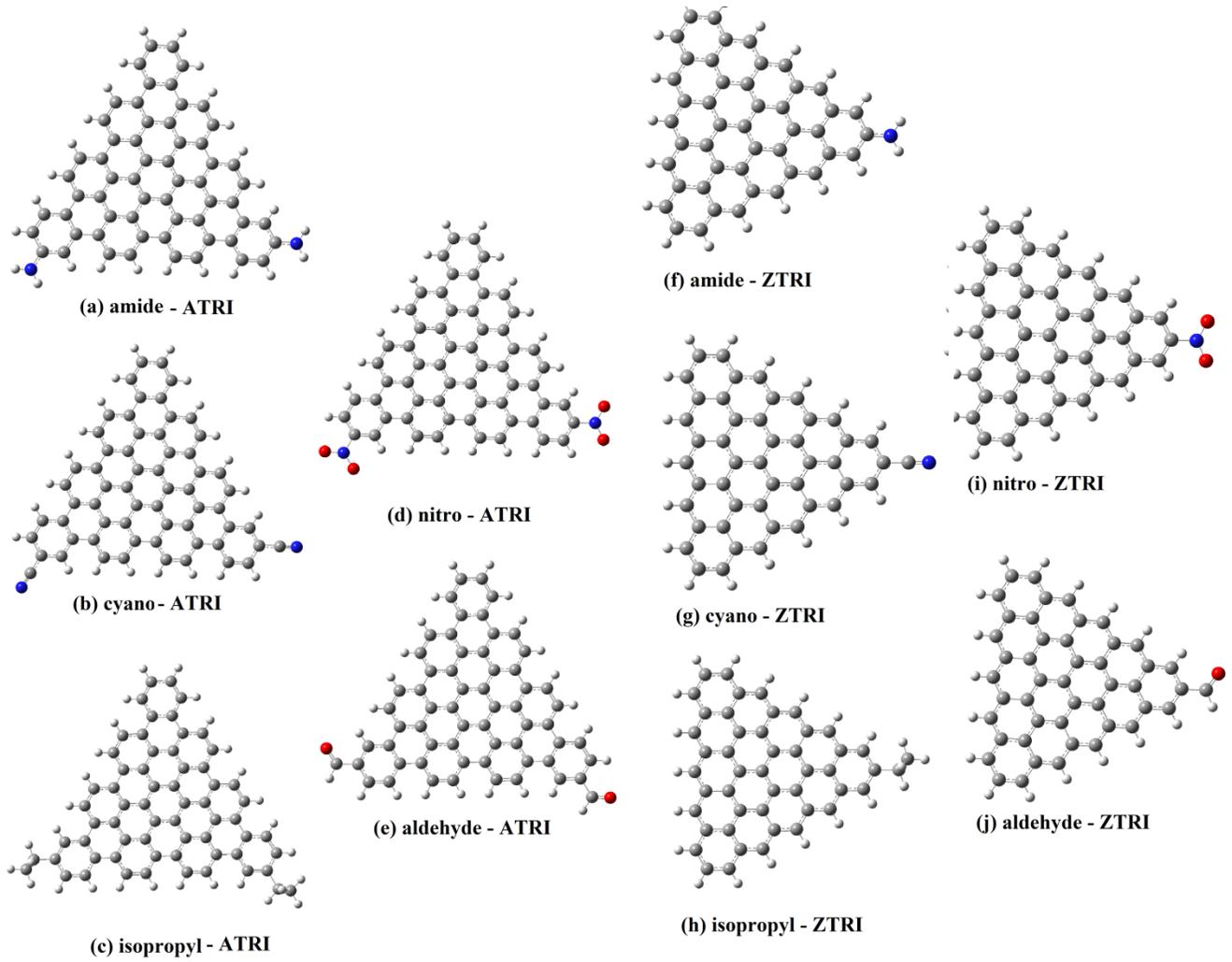

FIG. 2. The optimized structures for chemically functionalized ATRI (a-e) and ZTRI (f-j).

## C. ELECTRONIC DENSITY OF STATES AND ENERGY GAP

The density of states of the chemically functionalized hexagonal and triangular GQDs is calculated in term of Gaussian function $\frac{1}{\sqrt{2\pi}\alpha}\exp\left[\frac{(\varepsilon-\varepsilon_i)^2}{2\alpha^2}\right]$ that represent the energy levels with energy broadening $\alpha = 0.1$ eV. We consider a fixed Fermi level by setting $E_F = (E_{HOMO}+E_{LUMO})/2$ where $E_{HOMO}$ is the energy of highest occupied molecular orbital (HOMO) and $E_{LUMO}$ is the energy of the lowest unoccupied molecular orbital (LUMO). The electronic density of states (DOS) of AHEX and ZTRI are shown in FIG. 3 (a-e) and FIG. 3 (f-j), respectively. The density of states for ZHEX or ATRI has energy peaks, with wide energy gap between HOMO and LUMO, that is similar to that of AHEX thus it is not shown here. An obvious deference between DOS of the AHEX and ZTRI, seen from FIG. 3, is the largest energy gap observed in AHEX while the smallest gap is observed in ZTRI, namely in AHEX Eg ~ 3.4 eV and in ZTRI Eg ~ 0.3 eV. This large energy gap can be decreased to less than half its value by saturating edge atoms with oxygen instead of hydrogen, for O-AHEX the energy gap decreases to 1.45 eV.

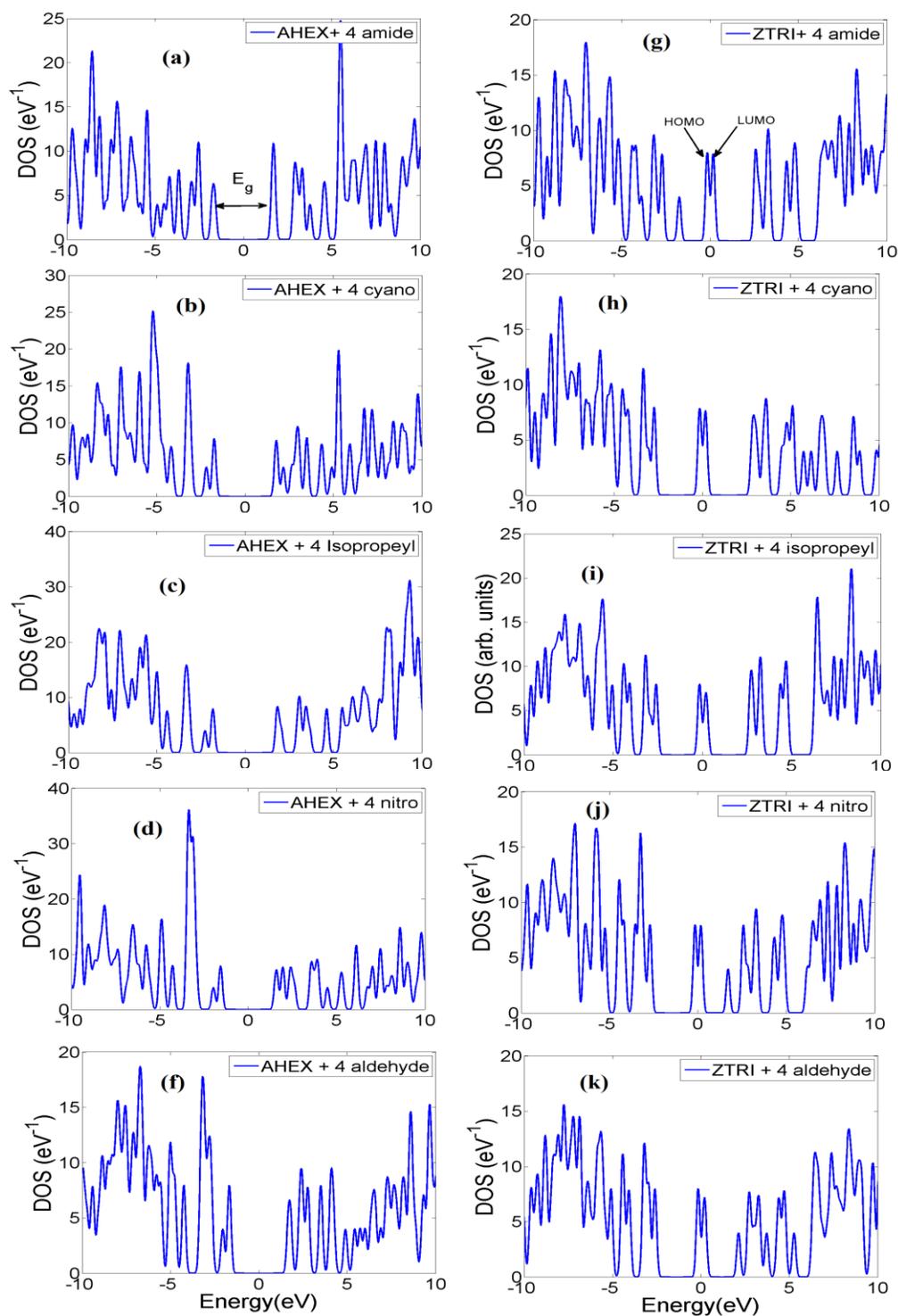

FIG. 3. DOS of the chemically functionalized AHEX (a-f) and ZTRI (g-k) clusters.

In FIG. 4 we show the distribution of HOMO and LUMO orbitals of AHEX and ZTRI clusters modified by different chemical groups. The HOMO and LUMO orbitals in AHEX are not localized and distributed over the surface of the cluster.

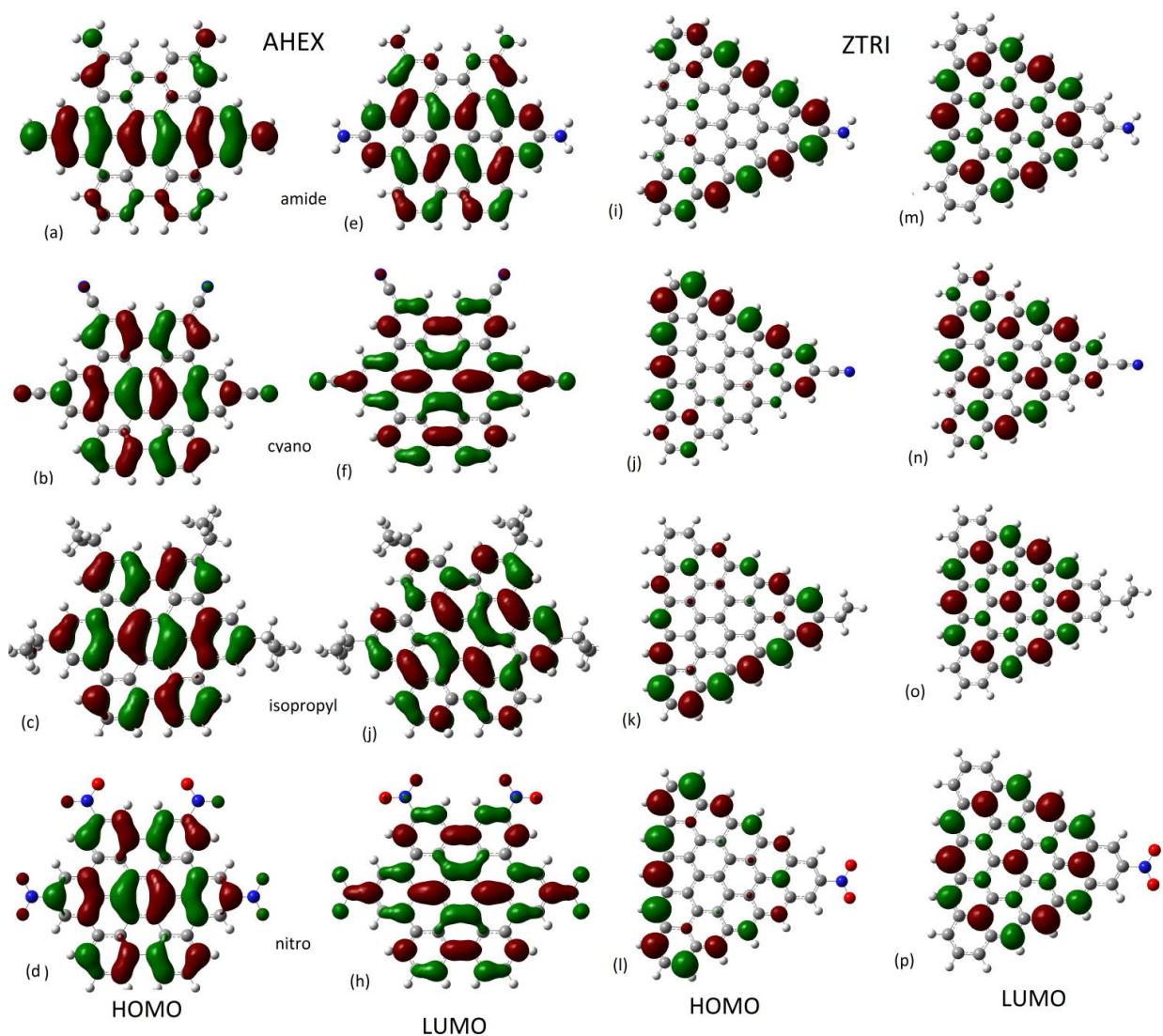

FIG. 4. The electron density distribution is shown for HOMO and LUMO of AHEX (a-h) and ZTRI (i-p) clusters functionalized by different groups.

While in ZTRI the HOMO is localized at the zigzag edges and the LUMO distributes over the surface of the cluster. The effect of chemical modification, see FIG. 4, is just slight redistribution of the HOMO and LUMO states but the general behavior of being localized at the edges or delocalized is not affected. Moreover in ZTRI the LUMO and HOMO states are localized to single atoms forming anti-bonding pi orbitals because the orbitals change color or sign on nearest atoms at the zigzag edges or surface, FIG. 4 (i-p). This type of bonding implies that these electrons are weakly bonded to the graphene cluster. This peculiar effect occurs only for ZTRI and is due to the fact that the HOMO and LUMO are edge states localized nearby the Fermi level. According to the tight binding model, these states are called zero energy states [21-24] and their number for ZTRI (C atoms =46) is 4 states. Therefore, the HOMO/LUMO shapes of sphere- like shape that is localized on single atoms will be observed only for HOMO, HOMO - 1, LUMO, and LUMO + 1.

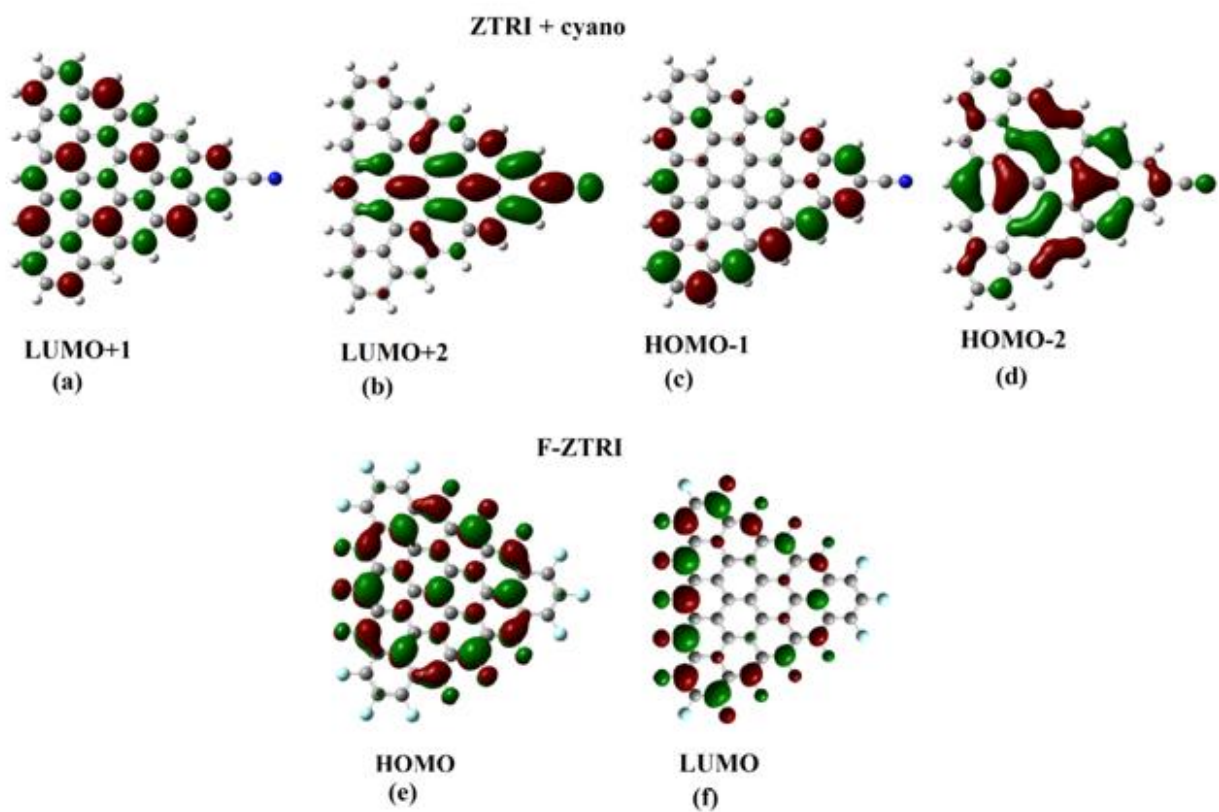

FIG. 5. (a-d) the distribution of LUMO + 1, LUMO+2, HOMO -1, and HOMO-2 orbitals in ZTRI +cyano. (e)/(f) shows HOMO/LUMO orbitals distribution for ZTRI passivated with fluorine.

To justify this explanation we plotted FIG. 5 showing the LUMO + 2, LUMO + 1, HOMO - 1, and HOMO - 2 for ZTRI + cyano. In addition to the HOMO/LUMO in FIG. 4 (j, n), it is clearly seen from FIG. 5 (a-d) that the distribution of the bulk HOMO - 2/LUMO + 2 orbitals is greatly different from that of the previously described edge states. Saturation of ZTRI cluster with hydrogen (H) open a very small energy gap between these 4 states and almost have no change on the HOMO and LUMO orbitals. In order to manipulate this states and the energy gap between them, we use other elements such as F to replace the H atoms. Passivation with fluorine increases the energy gap (from 0.27 in H-ZTRI to 0.34 eV in F-ZTRI) and considerably changes to the HOMO and LUMO orbitals distribution.

TABLE II. Electronic energy gap (ΔE) and total dipole moment (TDM) of GQDs attached to different functional groups.

| | Structure | ΔE (eV) | TDM (D) |
|---|---|---|---|
| ATRI | ATRI | 3.269 | 0.001 |
| | 2 amid | 3.07 | 3.9485 |
| | 2cyano | 3.19 | 6.0827 |
| | 2isopropeyl | 3.25 | 0.78 |
| | 2nitro | 2.83 | 6.83 |
| | 2aldehyde | 3.11 | 1.62 |
| ZTRI | ZTRI | 0.27 | 0.0035 |
| | amid | 0.31 | 3.75 |
| | cyano | 0.27 | 7.29 |
| | isopropyl | 0.31 | 1.69 |
| | nitro | 0.30 | 8.72 |
| | aldehyde | 0.31 | 5.89 |
| AHEX | AHEX | 3.668 | 0.000 |
| | 4amid | 3.27 | 5.46 |
| | 4cyano | 3.47 | 9.29 |
| | 4isopropeyl | 3.63 | 1.04 |
| | 4nitro | 3.09 | 9.99 |
| | 4aldehyde | 3.34 | 10.26 |
| ZHEX | ZHEX | 2.886 | 0.000 |
| | 5amid | 2.55 | 4.34 |
| | 5cyano | 2.72 | 8.19 |
| | 5isopropeyl | 2.84 | 1.07 |
| | 5nitro | 2.40 | 8.82 |
| | 5aldehyde | 2.57 | 8.77 |

Due to the high electronegativity of F with respect to H, a significant charge transfer is observed and the HOMO/LUMO states are now distributed over F atoms in addition to the ZTRI cluster as shown by FIG. 5 (e) and (f). Formation of the strong pi bonding between pi electrons on the edges is also observed in HOMO.

## C. TUNING THE TOTAL DIPOLE MOMENT

As seen from TABLE II, the effect of chemical modification on the energy gap is a slight decrease in AHEX and very small increase in ZTRI. In deep contrast, its effect on the dipole moment is considerably higher and depends on the attached group. We have obtained giant values of the total dipole moment (TDM) by chemical functionalization of different GQDs. The largest value, in TABLE II, is 10.3 (D) for AHEX + 4 aldehyde groups and the smallest value equals 0.78 (D) for ATRI + 2 isopropyl. Therefore, chemical functionalization in addition to the GQDs shape and edge morphology provide significant tunability of the TDM.

In order to study the origin of the wide range of the total dipole moment values, TABLE II, we plotted FIG. 6 to show the charge distribution on armchair-hexagonal GQDs (AHEX) without (a) and with attachment (b-f) of chemical groups.

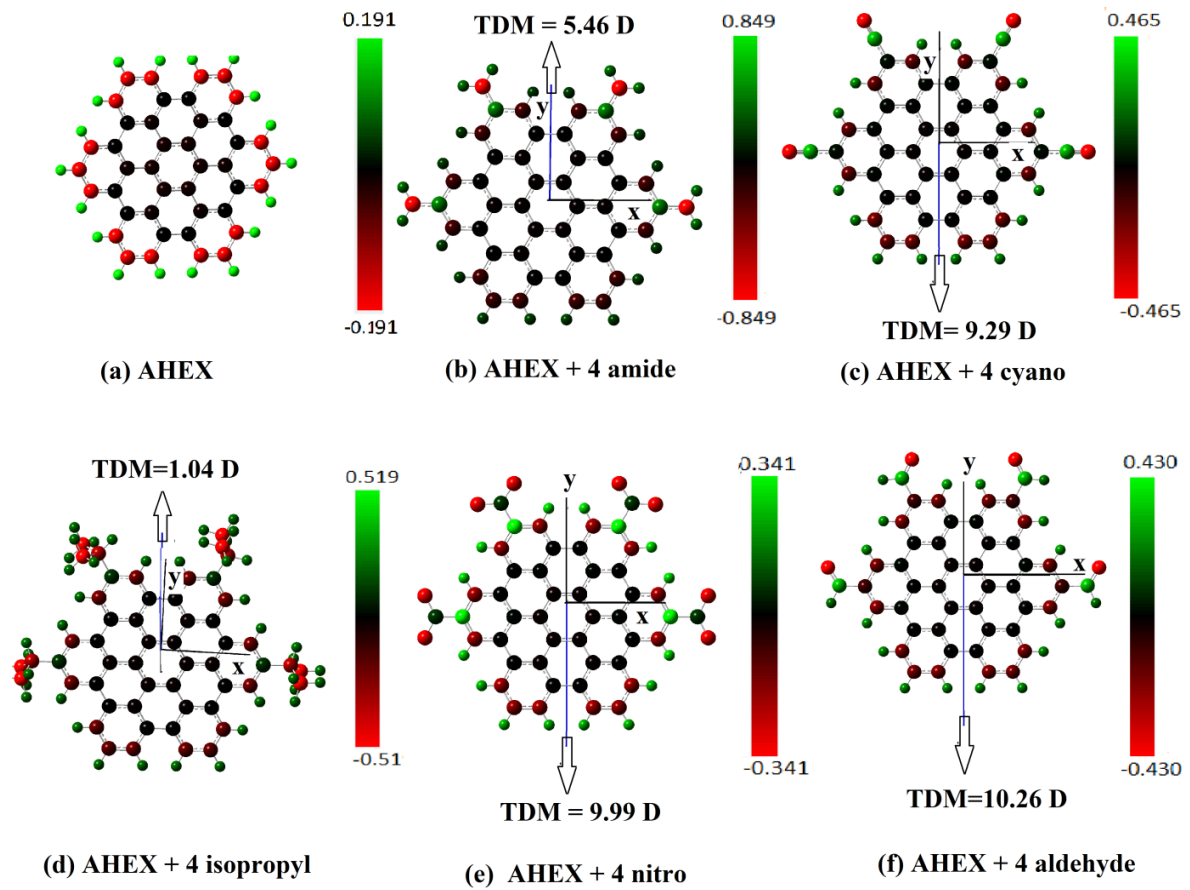

FIG. 6. Charge distribution of AHEX without (a) and with attachment to different groups (b-f).

For the non functionalized cluster (FIG. 6 a), as a result of its symmetry, the local dipole moments cancel each other providing a zero net dipole moment. Attachment of functional groups, according to the arrangement in FIG. 6, generates differences between local dipoles that give a nonzero value of TDM. The value of the TDM depends on the elements forming the group and its orientation in the 3D space. For example, the C≡N in cyano generates higher TDM than that by the C=O in amide. The configuration of attached groups leads to an overall increase or decrease of the TDM. For simplifying the discussion, only AHEX+ 4 cyano group will be considered to study the effect of the group orientation. As can be seen from FIG. 6 c, the x-component of the dipole moment cancel each other and the TDM are due to the y-component and z-component dipoles. The dipole moment in z-direction originated due to the buckling that takes place after the functionalization. However it adds only a negligible quantity to the TDM, for AHEX+ 4 cyano the dipole moment in z-direction is 0.0007 (D).

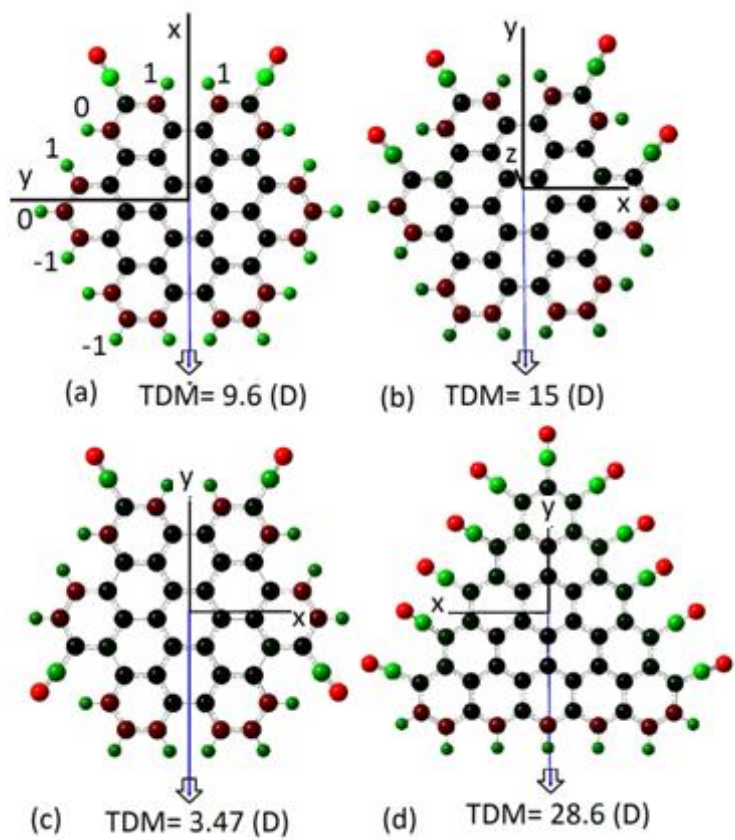

FIG. 7. The charge distribution and total dipole moment for selected AHEX and ZTRI attached to cyano groups.

Therefore almost all the value of TDM comes from the net dipole in the y-direction. This description is valid for all attached groups shown in FIG. 6 except for the isopropyl group the direction of TDM is little shifted from the y-direction. The geometric shape of the isopropyl group gives a nonzero value of dipole moment in x-direction which results in the small shifting from the y-direction. According to the previous discussion one can conclude that cyano groups to the left hand side (LHS) and to right hand side (RHS), FIG. 6 c, have almost no effect on the value of the TDM. However shifting them up or down to the nearest edge atoms can increase or decrease the TDM, respectively. The different positions to which these two cyano groups can be attached are shown in FIG. 7a. The attachment to position 0 adds negligible value to the TDM, to position 1 increases the TDM, and to position -1 decreases the TDM. This explanation is confirmed by calculations on AHEX+ 4 cyano cluster by (a) removing the two LHS and RHS groups, shifting them up and down. When removing the LHS and RHS cyano groups, the resultant TDM = 9.6 (D). This value is slightly higher than 9.29 (D) in TABLE II because the two LHS and RHS cyano groups (FIG. 7 c) are not perfectly aligned in x-direction. Shifting them up increase the TDM to14.99 (D) and when shifted down the TDM decreases to 3.47 (D). Moreover, when shifted to the positions -1 at the corners of the cluster, the local diploes cancel each other with negligible net TDM equals 0.001 (D).

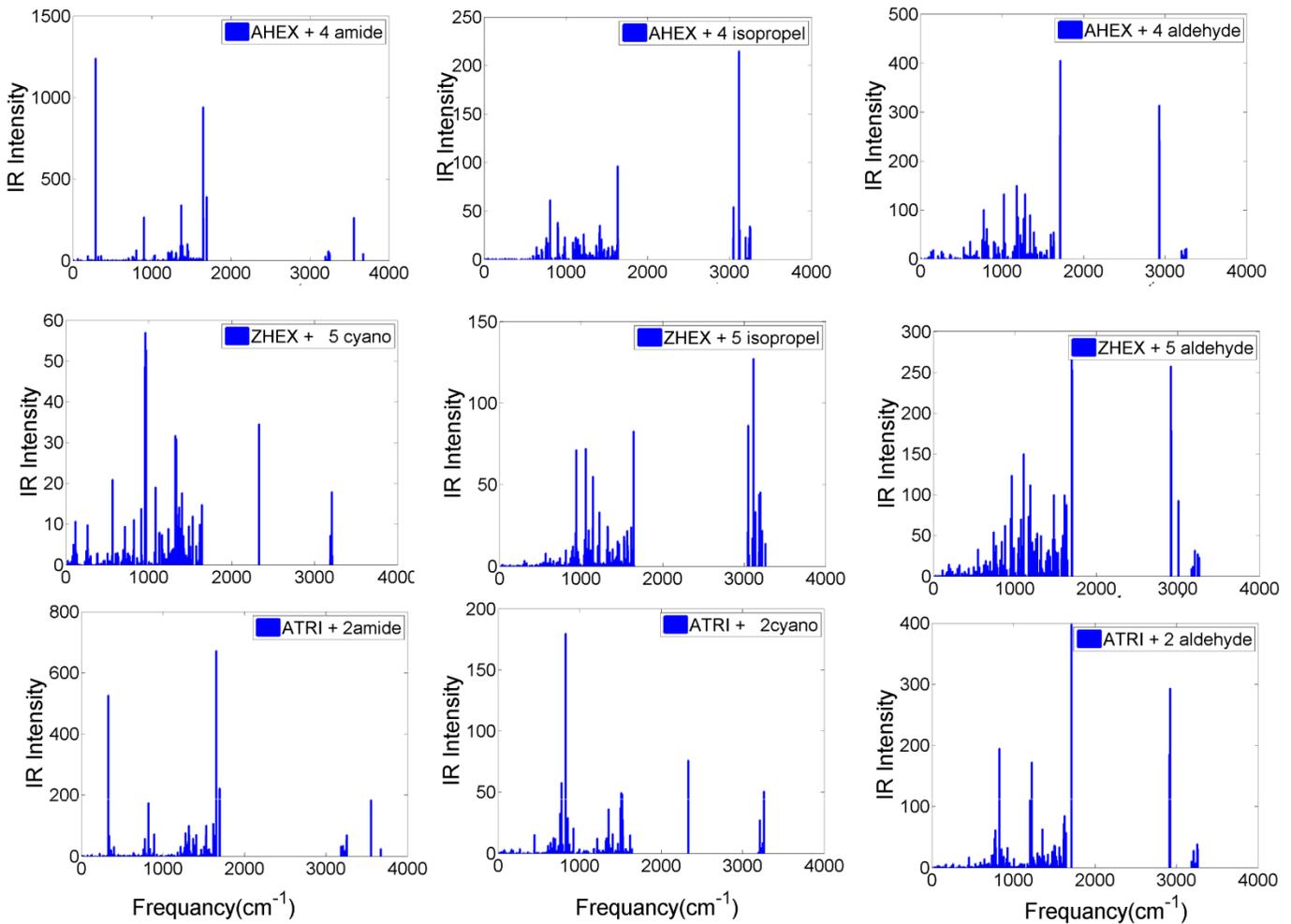

FIG. 8. The IR spectra of GQDs with various groups attached.

This idea can be applied to all other clusters, for example attaching cyano groups to two edges of the ZTRI increases (as in FIG. 7d) the TDM to 28.6 (D) but any further attachment will decrease the TDM. In conclusion, The TDM can be significantly tuned by changing the site of the attached groups in the same graphene cluster.

## D. INFRARED SPECTRA

Figures 8 and 9 show the infrared spectra for some selected chemically modified GQDs. The obtained real (positive) frequencies in all the clusters imply the stability of the studied GQDs.

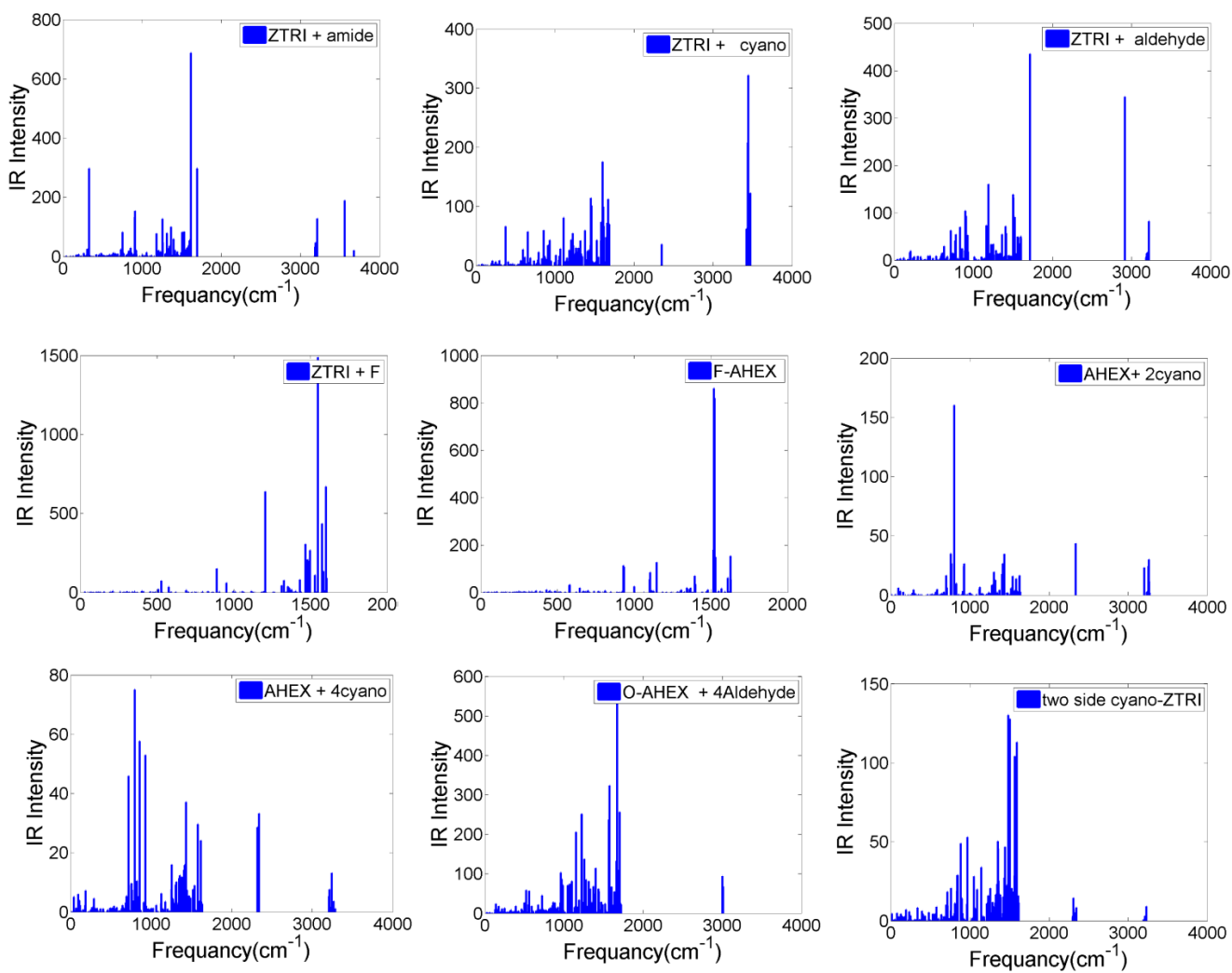

FIG. 9. The IR spectra of ZTR and AHEX attached to different groups and elements.

## IV.   CONCLUSION

In summary, the structure stability and electronic properties of edge functionalized triangular (ATRI and ZTRI) and hexagonal (AHEX and ZHEX) graphene quantum dots with zigzag and armchair termination have been studied using density functional theory. The calculations show that all the structures are stable with large energy gap in AHEX, ATRI, and ZHEX while a small energy gap is observed in ZTRI. This large energy gap can be significantly decreased by clusters passivation with oxygen, namely the energy gap decrease to less than half its value in AHEX when passivated with oxygen instead of hydrogen. The peculiar HOMO/LUMO orbitals and the small energy gap in ZTRI can also be manipulated by passivation with elements having high electronegativity such as fluorine.

The total dipole moment (TDM) of the selected cluster can be greatly tuned by chemical functionalization. Huge range of TDM values are obtained starting from zero (D) for ZHEX and AHEX to ~28 (D) in cyano-ZTRI. The TDM depends on: (a) the geometrical shape and edge termination, (b) the attached group, and (c) the position of the attached group. With respect to geometrical shape and attached group, the highest TDM was found in AHEX + 4 aldehyde because the local dipoles in negative y-direction generated by the four attached groups gather to increase total dipole in that direction. Positions of the attached groups have a strong effect on the TDM, in AHEX + 4 cyano the TDM can be significantly tuned by changing the position of cyano groups with respect to each other. In addition to the positive binding energy, the calculated frequencies are positive which insure that the chemically modified graphene quantum dots are thermodynamically stable. The tunable energy gap and the TDM in stable graphene clusters provided here make them ideal applicants for various applications such as transistors and biosensors.

## ACKNOWLEDGEMENTS

The authors are very grateful to V. A. Saroka for a careful reading and discussion of the manuscript and the suggested corrections.